# Molecular-absorption-induced thermal bistablity in PECVD silicon nitride microring resonators

Tingyi Gu,[1,*] Mingbin Yu,[2] Dim-Lee Kwong,[2] and Chee Wei Wong[1,*]

1Optical Nanostructures Laboratory, Columbia University, New York, NY 10027, USA

2The Institute of Microelectronics, 11 Science Park Road, Singapore Science Park II Singapore 117685, Singapore

*tg2342@columbia.edu, cww2104@columbia.edu

**Abstract:** The wavelength selective linear absorption in communication C-band is investigated in CMOS-processed PECVD silicon nitride rings. In the overcoupled region, the linear absorption loss lowers the on-resonance transmission of a ring resonator and increases its overall quality factor. Both the linear absorption and ring quality factor are maximized near 1520 nm. The direct heating by phonon absorption leads to thermal optical bistable switching in PECVD silicon nitride based microring resonators. We calibrate the linear absorption rate in the microring resonator by measuring its transmission lineshape at different laser power levels, consistent with coupled mode theory calculations.

1. Introduction

Interplay of free carrier and thermal nonlinearities in high quality factor ($Q$) small mode volume ($V$) optical resonators gives rise to various phenomena including self-pulsation and bistable switching [1-4]. Plasma-enhanced chemical vapor deposition (PECVD) produced materials, such as silicon nitride ($SiN_x$) and hydrogenated amorphous silicon (a-Si:H), are promising for the CMOS-compatible platform for low loss waveguide and resonators [5]. Along with experimentally demonstrated thermal and Kerr nonlinearities in PECVD grown devices [6-8], few works were found on studying the linear absorption mechanism leading to the thermal nonlinearities, where the electronic transitions related absorptions are weak in the wide bandgap materials. Different from electronic transitions, molecular/phonon absorption turns the optical energy directly to atomic lattice vibration and heat, represented by wavelength selective linear loss in waveguide (with bandwidth tens of meV) and thermal bistability in resonators [9-10]. In PECVD grown SiN$_x$ waveguides, the secondary harmonics of hydrogen-bond absorb the near-infrared (near-IR) light ~0.816 eV [11]. Near the absorption peak at 1520 nm, the thermal heating induced nonlinearity effectively shifts the cavity resonance [10] and clear bistable switching is shown in the ring with quality factor versus mode volume ratio $(Q/V)=10^{15}$ cm$^{-3}$. The resonator induced Purcell factor is ~$10^7$ for light-mater interaction at resonant wavelength [12]. The close investigation of the passive absorption process in resonators is useful for developing low power all optical bistable switch [13-15] and chemical sensors in infrared (IR) or mid-IR range [16]. Inversely, well understanding of the molecular absorption related linear and nonlinear response in resonators would provide useful information for optimizing their active functionalities, such as parametric oscillation, Raman amplification, cascaded four-wave mixing, where minimizing the linear absorption is a fundamental step towards low power and stable operation [17-25].

## 2. Device fabrication

SiN$_x$ films are grown by low-frequency low-temperature (350°C) PECVD (follow recipe No. 5 in [26]). The 80 sccm: 4000 sccm SiH$_4$:N$_2$ ratio and the 400W RF power control the deposition rate at 22.8A/s. PECVD silicon nitride has a Si/N ratio from 0.8 to 1, a density of 2.5 to 2.8g/cm$^3$, and refractive index of ~ 2.0 at 1550 nm wavelengths. The optical bandgap is estimated to be 3 to 4 eV [27]. We probe the components of silicon nitride film by the transmission Fourier-transform infrared (FTIR) methods.

With the low-temperature growth, low interlayer stress allows the SiN$_x$ deposition thickness up to 0.65 μm (Fig. 1(a)). Deep-UV lithography defines the 1 μm waveguide and ring widths with highly repeatable geometry, followed-by an optimized dry etch with high aspect ratios. The high aspect ratio allows good extinction between TE and TM polarizations (Fig. 1(b)). The optical properties of the waveguide, including refractive index and propagation losses can be optimized by controlling the plasma frequency, precursor gas ratio, and thermal annealing [26, 28]. The characteristic absorption wavelength at 1520 nm arises from the superposition of molecular rotations and vibrations. We verified the defects states absorption by FTIR spectroscopy (Fig. 2(a) and Fig. 2(b)). The small band around 485 cm$^{-1}$ is associated with Si breathing vibrations [29]. The absorption band at 863 cm$^{-1}$ is due to stretching vibration of Si-N bonds [30]. The bending vibration of Si-N bonds in the alpha-modification of crystalline silicon nitride is at 910 cm$^{-1}$ [31-32]. The bands located around 3356 and 1155 cm$^{-1}$ correspond to stretching and bending vibrations of N-H bonds. The Si-H bonds stretching mode is at 2205 cm$^{-1}$. Phonon modes beyond 4000cm$^{-1}$ could be a mix of high order phonon modes and the defects modes in the composite [11]. We performed the post-fabrication annealing at 800-1000$^{o}$C to reduce the molecular bonds, but annealing changes the cladding/core material morphology and led to extra scattering loss.

## 3. Measurement

Continuous-wave (CW) light generated from tunable laser (AQ4321) is sent onto chip through polarization controller and lensed fiber. An automatic power control circuit built in Ando laser AQ4321 maintains the optical output stability within +/- 0.01 (0.05) dB or less in 5 minutes (1 hour). With integrated spot size converter, the total fiber-chip-fiber loss is reduced to 14 dB. The output light is collected by both a slow power meter and a high-speed near-IR photoreceiver (New Focus Model 1554B, DC to 12 GHz bandwidth). The fast photoreceiver is connected to the digital phosphor oscilloscope (Tektronix TDS 7404, DC to 4 GHz bandwidth).

## 4. Linear absorption dependent quality factors in the over-coupled region

By scanning the tunable laser from 1480 nm to 1560 nm, we measured the power transmission for a range of waveguide lengths from 25 mm to 44 mm with 5 mm discrete steps. The fitted transmitted power versus waveguide length with linear model gives a propagation loss of 4.3 dB/cm at 1550 nm, and a -13.5 dB insertion loss from the two facets of waveguides. However, we observed the total loss is strongly wavelength dependent, arising largely from molecular-bond absorption in the bulk nitride [11]. The propagation loss has an absorption peak at 1520 nm with 37.2 nm (20 meV) full-width half-maximum (Fig. 2(b)). The maximum loss is 6.8dB/cm at 1520 nm and reduces to 2.3dB/cm when the wavelength is detuned 25 nm away from the absorption peak. The propagation scattering loss due to scattering is mostly wavelength independent and about 2 dB/cm.

The wavelength-dependent linear absorption (Fig. 2(b)) is presented by the dip in waveguide transmission spectrum (black solid curve in Fig. 2(c)). The propagation loss composes of waveguide sidewall roughness scattering and material absorption in near-IR. The material absorption near 1520 nm is mostly induced by middle band defects absorption. The coupling matrix gives the linear absorption dependent on-resonance transmission [33-34] as:

$$T_{res} = [\frac{t-\alpha(\lambda_{res})}{1-\alpha(\lambda_{res})t}]^2 \quad (1)$$

here $t$ is the field transmittance between ring and waveguide. $\alpha(\lambda)=exp(-\alpha_L(\lambda)L/2)$ is the round-trip field transmission of the ring cavity with circumference $L$ ($2\pi\times40\mu m$), and intrinsic propagation coefficient $\alpha_L$. $\alpha_L$ is wavelength dependent for the PECVD grown silicon nitride waveguide, and derived from our waveguide transmission measurements. We plot the on-resonance transmission ($T_{res}$) as a function of the field transmittance $\alpha$ for the resonances (inset of Fig. 2(c)). As the round-trip transmittance (1- $\alpha(\lambda)$) decreases from 0.99 to 0.9, the on-resonance transmittance drops from 0.75 to 0.2 in the overcoupling region (red curve in the inset of Fig. 2(c)). The on-resonance transmission is minimized at the critical coupling point, where the round-trip transmittance has the same value as ring-waveguide field transmittance ($t = 0.89$). The trend is opposite as the linear absorption keeps increasing into the undercoupling region. The on-resonance transmission increases with the linear loss in the under-coupling region (blue dashed line in the inset of Fig. 2(c)), where the quality factor decreases with linear loss in waveguide as usually observed.

With the wavelength-dependent linear loss, the resonator quality factor over the whole spectrum is plotted in Fig. 2(d). The increasing quality factor with higher linear propagation loss is due to the approach into critical coupling regime between the ring and the waveguide. The total $Q$ factor (experimentally measured as $\lambda_{res}/\Delta\lambda$, where $\lambda_{res}$ and $\Delta\lambda$ are obtained by applying Lorenzian fit to individual resonance) of a microring resonator coupled to a single waveguide can be expressed as [35-36]:

$$Q_{total}(\alpha) = \frac{2\pi n_{eff}}{\lambda_0} \frac{L}{2\arccos[2-(0.5/t')\exp(-\alpha_L(\lambda)L/2)-0.5t'\exp(\alpha_L L(\lambda)/2)]} \quad (2)$$

where $n_{eff}$=1.6 is the effective refractive index of silicon nitride waveguide. $\lambda_0$ is the resonance wavelength of the ring, $t'$ is the field transmission coefficient between the ring and waveguide. By fitting the model to the measured data, we obtained the field transmission coefficient $t' = 0.93$. It is noted that the trend is inverse with $t' = 1$. The maximized linear absorption and ring quality factor near 1520 nm form an ideal condition for investigating the optical nonlinearity from the light matter interaction. With the wavelength dependent linear loss (black line in Fig. 2(c)), equation 2 predicts the correspondent quality factor (circles in Fig. 2(d)), which is comparable to experimental measurement (crosses in Fig. 2(d)).

We measured three rings with radii of ~20, 40 and 70 μm, with loaded, intrinsic quality factor and FSR respectively of 24,500, 49,000, 8.7nm, 69,600, 175,000, 4.4 nm and 77,300, 244,000, 2.9 nm at 1550 nm. The 40 μm radius ring is presented here for nonlinearity investigation due to its highest $Q/V$ ratio. The loaded and intrinsic factors are obtained by careful coupled-mode-theory curve fitting at different power levels. To show the linear and nonlinear effect of absorption rate to resonator lineshape, we compared the linear and nonlinear response of the ring with different material absorption lifetime in ring resonator ($\tau_{lin}$).

## 5. Thermal induced bistable switching

Comparing the resonances at different wavelengths, the linear loss modifies the ring-waveguide coupling coefficient. The linear loss enlarges extinction ratio in the overcoupling region. The steepened transmission spectrum shows higher the quality factor. With fixed ring-waveguide coupling coefficient, we model the nonlinear cavity transmissions with time-domain nonlinear coupled mode theory for the intracavity photon and temperature dynamics [37]:

$$\frac{da}{dt} = (i(\omega_L - \omega_0 - \Delta\omega_T) - \frac{1}{2\tau_t})a + \kappa\sqrt{P_{in}} \qquad (3\text{-}1)$$

$$\frac{d\Delta T}{dt} = \frac{R_{th}}{\tau_{th}\tau_{lin}}|a|^2 + \frac{\Delta T}{\tau_{th}} \qquad (3\text{-}2)$$

where $a$ is the amplitude of resonance mode; $\Delta T$ is the cavity temperature shift. $P_{in}$ is the power carried by incident CW laser wave. $\kappa$ is the coupling coefficient between waveguide and cavity, adjusted by linear absorption in waveguide. $\omega_L$-$\omega_0$ is the detuning between the laser frequency ($\omega_L$) and cold cavity resonance ($\omega_0$). The thermal induced dispersion is $\Delta\omega_T=\omega_0\Delta T(dn/dT)/n$. The total loss rate is $1/\tau_t = 1/\tau_{in}+1/\tau_v+1/\tau_{lin}$. $1/\tau_{in}$ and $1/\tau_v$ are the loss rates into waveguide (coupling loss) and into free space (scattering loss), ($1/\tau_{in/v} =\omega/Q_{in/v}$). The defect absorption leads to the linear absorption rate: $1/\tau_{lin}=c\alpha/n$. The linear absorption rate, as a product of defects density and absorption cross section ($\alpha=\sigma_{def}N_{def}$), is estimated to be 0.0017 cm$^{-1}$. In steady state condition here, Kerr dispersion is negligibly small compared to the thermal effect and thus not included here. The rest of linear and nonlinear parameters in silicon nitride ring are listed in the Table 1. It is noted that not all the parameters are independent in table I. The thermal relaxation time $\tau_{th,c}= R_{th}\times c_v\rho\times V$. The mode volume $V=2\pi rS$, where $S=0.1\mu m^2$ is FDTD calculated the mode area on the cross section of the waveguide with given dimension in the right inset of Fig. 1(a).

The side coupling loss rate of the ring ($\omega_0/Q_{in}/2$) is 72.5 GHz, and the linear absorption rate ($1/\tau_{lin}$) is 2.5 GHz (Fig. 3(a)). The coupling and intrinsic quality factors are 85,000 and 1,500,000 respectively. The transmission lineshape evolves from symmetric Lorentzian to unsymmetrical bistable one as the input power increases from 20 µW (Fig. 3(a)) to 156 µW (Fig. 3(b)). Good model-experiment correspondence at different input power levels is obtained by maintaining the same parameter space (Fig. 3(c)). The power dependent transmission spectrum for TM modes near the 1520nm resonance is similar to the TE modes, confirming the linear absorption related thermal nonlinearity is polarization insensitive. As the intensity built-up factor is proportional to $Q^2/V$, the switching energy dramatically reduces with linear absorption (inversely proportional to $\alpha\times Q^2/V$), and minimized to 1.38pJ near chemical bond absorption peak at 1520nm. For the ring resonance far away from the molecular absorption dip (1560nm), the transmission spectrums at different power levels (from 20 to 156µW) have the same symmetric Lorentzian lineshape, but with slightly shifted resonance wavelength.

The tuning efficiency increases to a peak value at 80 pm/mW for the resonances at 1520 nm wavelength (for both TE and TM polarizations), and drops to 10 pm/mW near 1610 nm resonance under optimized polarization and coupling control. Fig. 3(d) plots cavity resonance shift versus the input power. The straight linear relation implies negligible contribution from higher order nonlinear absorption such as two-photon absorption and free-carrier absorption related thermal nonlinearities. Near the absorption peak, the input/output transfer function shows clear bistability when the laser detuning ($\delta=(\lambda_L-\lambda_{res0})/(\Delta\lambda/2)$) is set at $\delta = 3.42$ and 3.72 (inset of Fig. 3(d)), where $\lambda_L$, $\lambda_{res0}$, and $\Delta\lambda$ are laser wavelength, cold cavity resonance and cavity bandwidth respectively.

The optical absorption in near 1520 nm is dominated by absorption from the N-H bonds. Through the model-measurement correspondence at different power levels, the linear absorption lifetime in the ring resonator is 400 ps (Fig. 3). Higher linear absorption (absorption rate) lowers the extinction ratio in linear region (dashed red line in Fig. 3(a)) and increases cavity resonance shift at nonlinear region (dashed red line in Fig. 3(b)). The transmission spectrum measurement at different power levels are shown in Fig. 3(c), with the parameter space given in Table 1.

Table 1. Fixed parameters used in the CMT model

| Parameter | Symbol (unit) | SiN |
|---|---|---|
| $SiN_x$ refractive index | $n$ | 2.03 |
| Radius of the ring | $r$ (μm) | 40 |
| Mode volume | $V$(μm$^3$) | 25.1 [FDTD] |
| Loaded $Q$ | $Q$ | 85,000 [CMT] |
| Intrinsic $Q$ | $Q_0$ | 1,500,000 [CMT] |
| Thermo-optic coeff. | $dn/dT$ (K$^{-1}$) | $2.6\times10^{-5}$ [38] |
| Scattering rate | $1/\tau_v$ (GHz) | 0.42 [Cal] |
| Linear absorption rate | $1/\tau_{lin}$ (GHz) | 2.5 [CMT] |
| Coupling rate | $1/\tau_{in}$ (GHz) | 72.5 [Cal] |
| Heat capacity | $C$ (J/K/kg) | 700 [39] |
| Specific heat | $c_v\rho$ (W/Km$^{-3}$) | $1.84\times10^6$ |
| Thermal resistance | $R_{th}$ (K/mW) | 17.5 [Cal] |
| Thermal relaxation time | $\tau_{th,c}$ (ns) | 808 [Cal] |

[CMT]: couple mode theory curve fitting; [Cal]: Derived value from other parameters; [FDTD]: Finite difference time domain method calculation.

We also performed two-wavelength bistable operation to achieve the all optical bistable switch functionality, with one pump laser power changes the refractive index through thermal-optic dispersion and another probe laser to separately monitor the resonance shift [14,40]. One cw pump laser is set on mode near 1540 nm and the wavelength of another probe laser is set near resonance of another adjacent mode. The power of the probe laser is below 22μW not to trigger nonlinear response. The output of the pump power is blocked by a 30 dB notch filter, and only the probe power is collected by the photodetector. We clearly observe transmission of probe laser decreases as nonlinearity shifts as the cavity resonance approaching probe laser wavelength. The probe laser transmission is minimized at pump power level shifting the resonance onto the probe laser wavelength. The output power of the probe laser increases as the pump power keep increasing and shifting the resonance away from the probe laser wavelength (Fig. 4(a)). The plot of the probe detuning versus the pump power dropped into the resonator gives the linear relation between cavity resonance shift efficiency at 55 pm/mW at 1540 nm (inset of Fig. 4(a)), consistent with the value ins Fig. 3(d) [41]. The related thermal dynamics is measured at both positive and negative pump laser-cavity detunings in Fig 4(b). We observed the first-order exponential decay of the probe transmission, when a step-input function (to 1 mW) is entered on the cw pump laser. The exponential fit gives thermal lifetime of ~150 μs for the nitride ring resonator system. The measured lifetime may be related to the heat dissipation to substrate rather than the core-cladding system ($\tau_{th,c}$) given in table I (also in [42]), and is consistent with other chemical bond absorption related thermal nonlinearities [16].

## 6. Conclusions

The wavelength selective molecular absorption modifies the linear absorption of the PECVD silicon nitride ring resonators, and led to thermal bistability. The absence of free carrier dispersion could improve the stability of thermal bistable switching. Three independent models are compared to experimental data for correlating the linear loss in waveguide to microring resonator behavoir. With fixed scattering loss, higher linear loss from material absorption steepens the extinction ratio in the overcoupling region, and enhances the empirical quality factor of the resonator. The phonon absorption, coupled with enhanced Q factors, leads to pico joule level bistable switching in the wide bandgap material based microring resonator. The power dependent nonlinear transmission spectrum is numerically described by the couple mode theory.

**Acknowledgements**

The authors acknowledge informative discussions with M. Voros and N. Brawand from Prof. J. Galli's group. T. Gu is grateful for technical discussion with A. Raja and Y. Li and FTIR measurement in Prof. T. F. Heinz' group. T. Gu thanks X. Zhang for device annealing. T. Gu appreciates discussions with Prof. R. M. Osgood Jr. and Prof. L. Brus. The authors acknowledge support from the Office of Naval Research on the nonlinear dynamics program under contract N00014-14-1-0041. This material is based upon work supported as part of the Center for Re-Defining Photovoltaic Efficiency through Molecule Scale Control, an Energy Frontier Research Center funded by the U.S. Department of Energy, Office of Science, Office of Basic Energy Sciences under Award Number DE-SC0001085.

**References*:*
1. X. Yang, C. Husko, C. W. Wong, M. Yu, and D. L. Kwong, "Observation of femtojoule optical bistability involving Fano resonances in high-$Q/V_m$ silicon photonic crystal nanocavities," Appl. Phys.Lett. **91**(5), 051113-051113 (2007).
2. C. Husko, A. De Rossi, S. Combrié, Q. V. Tran, F. Raineri, and C. W Wong,. "Ultrafast all-optical modulation in GaAs photonic crystal cavities," Appl. Phys. Lett. **94**, 021111 (2009).
3. J. Yang, T. Gu, J. Zheng, M. Yu, G.-Q. Lo, D.-L. Kwong, and C. W. Wong, Radio-frequency regenerative oscillations in monolithic high-Q/V heterostructured photonic crystal cavities, *Appl. Phys. Lett.* **104**, 061104 (2014).
4. T. Gu, N. Petrone, J. F. McMillan, A. van der Zande, M. Yu, G. Q. Lo, and C. W. Wong, "Regenerative oscillation and four-wave mixing in graphene optoelectronics," Nat. Photon. **6**(8), 554-559 (2012).
5. D. J. Moss, R. Morandotti, A.L. Gaeta, and M. Lipson, "New CMOS-compatible platforms based on silicon nitride and Hydex for nonlinear optics," Nat. Photon. **7**, 597-607 (2013).
6. K. Ikeda, R. E. Saperstein, N. Alic and Y. Fainman, "Thermal and Kerr nonlinear properties of plasma-deposited silicon nitride/silicon dioxide waveguides," Opt. Exp. **16**(17), 12987-12994 (2008).
7. J. S. Pelc, K. Rivoire, S. Vo, C. Santori, D. A. Fattal, and R. G. Beausoleil, "Picosecond all-optical switching in hydrogenated amorphous silicon microring resonators," Opt. Exp. **22** (4), 3797-3810 (2014).
8. A. Gondarenko, J. S. Levy, and M.. Lipson, "High confinement micron-scale silicon nitride high *Q* ring resonator," Opt. Exp. **17**(14), 11366-11370 (2009).
9. T. Carmon, L. Yang, and K. Vahala. "Dynamical thermal behavior and thermal self-stability of microcavities," Opt. Exp. **12**(20), 4742-4750 (2004).
10. H. Rokhsari, S. M. Spillane, and K. J. Vahala, "Loss characterization in microcavities using the thermal bistability effect," Appl. Phys. Lett. 85(15), 3029-3031 (2004).

11. C.H. Henry, R. F. Kazarinov, H. J.Lee, K. J. Orlowsky and L. E. Katz, "Low loss $Si_3N_4$-$SiO_2$ optical waveguides on Si," 26(13), 2621-2624 (1987).
12. J. Vučković, M. Lončar, H. Mabuchi, and A. Scherer, "Design of photonic crystal microcavities for cavity QED," Phys. Rev. E 65(1), 016608 (2001).
13. Z.-G. Zang and Y.-J.Zhang, "Low-switching power (<45 mW) optical bistability based on optical nonlinearity of ytterbium-doped fiber with a fiber Bragg grating pair," J. of Mod. Opt., **59**(2), 161-165 (2012).
14. M. Notomi, A. Shinya, S. Mitsugi, G. Kira, E. Kuramochi, and T. Tanabe, "Optical bistgable switching action of Si high-Q photonic-crystal nanocavities, " Opt. Exp. **13**(7), 2678-2687 (2005).
15. V. R. Almeida and M. Lipson, "Optical bistability on a silicon chip," Opt. Lett. **29**(20), 2387-2389 (2004).
16. M. W. Lee, C. Grillet, C. Monat, E. Mägi, S. Tomljenovic-Hanic, X. Gai, S. Madden, D.-Y. Choi,D. Bulla, B. Luther-Davies, and B. J. Eggleton, "Photosensitive and thermal nonlinear effects in chalcogenide photonic crystal cavities," Opt. Exp. **18**(25), 26695-26703 (2010).
17. J. S. Levy, A. Gondarenko, M. A. Foster, A. C. Turner-Foster, A. L. Gaeta, and M. Lipson, "CMOS-compatible multiple-wavelength oscillator for on-chip optical interconnects," Nat. Photon. **4**(1), 37-40 (2009).
18. S.-W. Huang, J. F. McMillan, J. Yang, A. Matsko, H. Zhou, M. Yu, D.-L. Kwong, L. Maleki, C. W. Wong, "Direct generation of 74-fs mode-locking from on-chip normal dispersion frequency combs," submitted under review (2014)
19. K.-Y. Wang, and A. C. Foster, "Ultralow power contiuous-wave frequency conversion in hydrogenated amorphous silicon waveguides," Opt. Lett. **37**(8), 1331-1333 (2012).
20. K. Narayanan, S.F. Preble, "Optical nonlinearities in hydrogenated-amorphous silicon waveguides," Opt. Lett. **18**(9), 8998-9005 (2010)
21. L. Razzari, D. Duchesne, M. Ferrera, R. Morandotti, S. Chu, B. E. Little, and D. J. Moss, "CMOS-compatible integrated optical hyper-parametric oscillator," Nat. Photon. **4**, 41-45 (2010).
22. S. M. Spillane, T. J. Kippenberg, and K. J. Vahala, "Ultralow-threshold Raman laser using a spherical dielectric microcavity," Nature **415**, 621–623 (2002).
23. N. Carlie, J. D. Musgraves, B. Zdyrko, I. Luzinov, J. Hu, V. Singh, A. Agarwal, L. C. Kimerling, A. Canciamilla, F. Morichetti, A. Melloni, and K. Richardson, "Integrated chalcogenide waveguide resonators for mid-IR sensing: leveraging material properties to meet fabrication challenges," Opt. Exp., **18**(25), 26728-26743 (2010).
24. Y. Zha, P. T. Lin, L. Kimerling, A. Agarwal, and C. B. Arnold, "Inverted-Rib Chalcogenide Waveguides by Solution Process," ACS Photonics, **1**(3) 153-157(2014).
25. K. Vu, K. Yan, Z. Jin, X. Gai, D.-Y. Choi, S. Debbarma, B. Luther-Davies, and S. Madden, "Hybrid waveguide from As2S3 and Er-doped TeO2 for lossless nonlinear optics," Opt. Lett. **38**(11), 1766-1768 (2013).
26. S. C. Mao, S. H. Tao, Y. L. Xu, X. W. Sun, M. B. Yu, G. Q. Lo, and D. L. Kwong, "Low propagation loss SiN optical waveguide prepared by optimal low-hydrogen module," Opt. Exp. **16**(25), 20809-20816 (2008).

27. S. V. Deshpande, E. Gulari, S. W. Brown, and S. C. Rand, "Optical properties of silicon nitride films deposited by hot filament chemical vapor deposition," J. Appl. Phys. **77**, 6534 (1995).
28. A. Gorin, A. Jaouad, E. Grondin, V. Aimez, and P. Charette, "Fabrication of silicon nitride waveguides for visible-light using PECVD: a study of the effect of plasma frequency on optical properties," Opt. Exp. **16**(18), 13509-13516 (2008).
29. D. V. Tsu, G. Lucovsky, and B. N. Davidson, "Effects of the nearest neighbors and the alloy matrix on SiH stretching vibrations in the amorphous $SiO_r$: H ($0< r< 2$) alloy system," Phys. Rev. B **40**(3), 1795 (1989).
30. J. Fandiño, A. Ortiz, L. Rodríguez-Fernandez, and J. C. Alonso, "Composition, structural, and electrical properties of fluorinated silicon-nitride thin films grown by remote plasma-enhanced chemical-vapor deposition from $SiF_4/NH_3$ mixtures," J. Vac. Sci. Tech. **22**, 570-577 (2004).
31. D. V. Tsu, and G. Lucovsky, "Silicon nitride and silicon diimide grown by remote plasma enhanced chemical vapor deposition," J. Vac. Sci. Tech. A: Vacuum, Surfaces, and Films **4**(3), 480-485 (1986).
32. N. Wada, S. A. Solin, J. Wong, and S. Prochazka, "Raman and IR absorption spectroscopic studies on α, β, and amorphous $Si_3N_4$," J. Non-Crystalline Solids **43**(1), 7-15 (1981).
33. A. Yariv. "Critical coupling and its control in optical waveguide-ring resonator systems," Photon. Tech. Lett., IEEE **14**(4), 483-485 (2002).
34. H. Y. Wen, O. Kuzucu, M. Fridman, A. L. Gaeta, L.-W. Luo, and M. Lipson. "All-optical control of an individual resonance in a silicon microresonator," Phys. Rev. Lett. **108**(22), 223907 (2012).
35. J. Niehusmann, A. Vörckel, P. H. Bolivar, T. Wahlbrink, W. Henschel, and H. Kurz, "Ultrahigh-quality-factor silicon-on-insulator microring resonator," Opt. Lett. **29**(24), 2861-2863 (2004).
36. A. Vorckel, M. Monster, W. Henschel, P. H. Bolivar, and H. Kurz, "Asymmetrically coupled silicon-on-insulator microring resonators for compact add-drop multiplexers," IEEE Photon. Tech. Lett. **15**, 921-923 (2003).
37. H. A. Haus, Waves and Fields in Optoelectronics, Prentice-Hall, Englewood Cliffs, NJ (1984).
38. G. S. Wiederhecker, L. Chen, A. Gondarenko, and M. Lipson, "Controlling photonic structures using optical forces," Nature **462**(7273), 633-636 (2009).
39. C. Baker, S. Sebastian, D. Parrain, S. Ducci, G. Leo, E. M. Weig, and I. Favero, "Optical instability and self-pulsing in silicon nitride whispering gallery resonators," Opt. Exp. **20**(27), 29076-29089 (2012).
40. A. Arbabi and L. L. Goddard, "Measurements of the refractive indices and thermo-optic coefficients of $Si_3N_4$ and $SiO_x$ using microring resonances," Opt. Lett. **38** (19), 3878-3881 (2013).
41. P. E. Barclay, K. Srinivasan, and O. Painter, "Nonlinear response of silicon photonic crystal micro-resonators excited via an integrated waveguide and fiber taper," Opt. Exp. **13**(3), 801 (2005).
42. A. Arbabi, and L. L. Goddard, "Dynamics of self-heating in microring resonators," Photon. J., IEEE 4, 1702-1711 (2012).

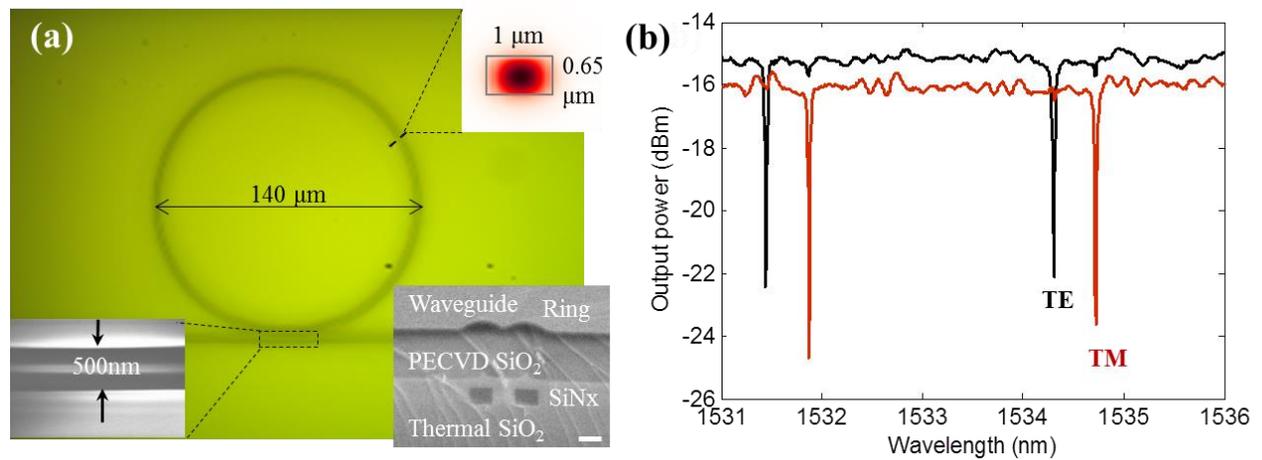

Fig. 1. Structure and linear optical properties of the device (a) Silicon nitride device layout. Optical image of top view of the ring, where the dashed line shows the cleaved position for the SEM image. Inset (up right): Cross section and the optical profile of the TE mode. Inset (bottom left): SEM image of the ring-waveguide coupling part. Inset (bottom right): 650nm PECVD silicon nitride is sandwiched between the PECVD silicon oxide up cladding layer and the thermal oxide lower cladding layer. Scale bar: 1μm. (b) Output spectrum of TE and TM polarized input with 0dBm input power.

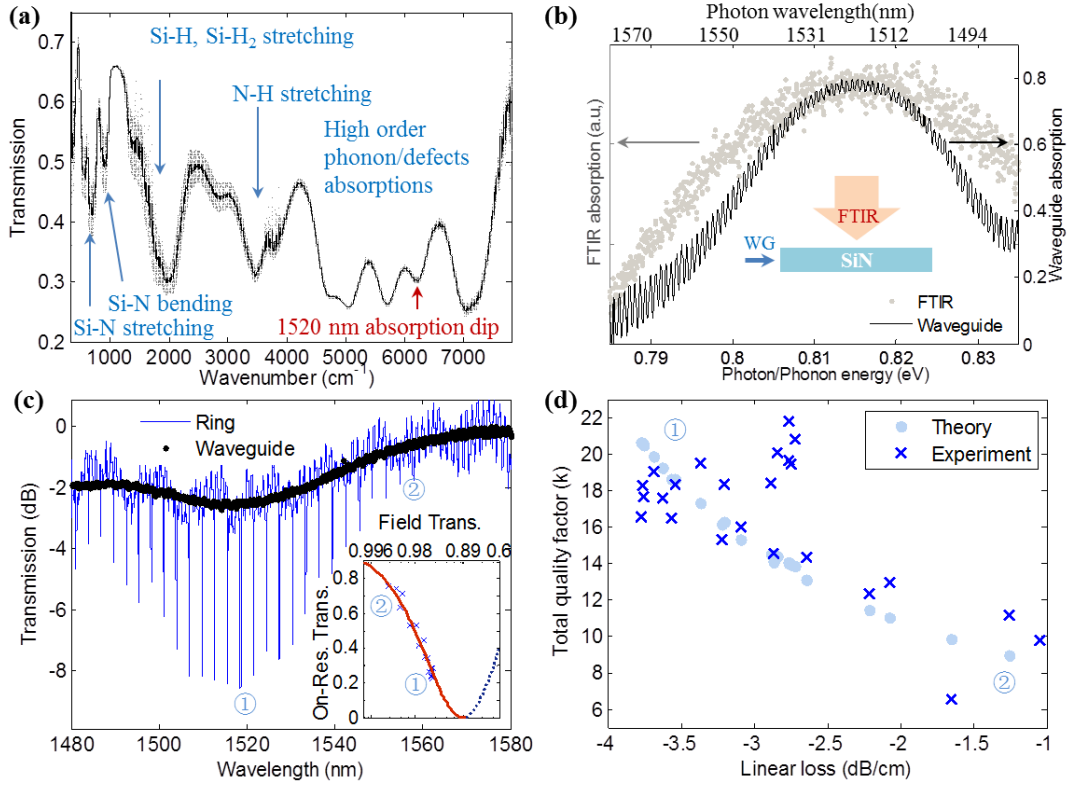

Fig. 2. Wavelength selective absorption and ring quality factors (a) Linear absorption of the PECVD grown silicon nitride thin film, in the range of mid- and near-IR. (b) FTIR measured absorption versus phonon energy of PECVD silicon nitride thin film (gray dots) and the absorption of a 25 mm long SiN waveguide versus photon energy from tunable laser (black solid line). Wavelength dependent linear propagation loss is maximized near 1520 nm. The absorption peak is at 0.816 eV with FWHM of 20 meV. Inset: schematics of incident light direction for waveguide and FTIR measurement. (c) Normalized transmission of ring resonator of 70 μm radius (blue) and 6.7 mm long waveguide (black). Inset: On-resonance transmission versus intracavity field transmission $(1-\alpha(\lambda)L)$. The blue crosses are experimental data. Red solid line and blue dashed line are theoretical predictions for over-coupled and under-coupled regions respectively. (d) Linear loss dependent total quality factors. Experimental results are directly derived from fitting the ring resonances in (c), and theoretical predictions are given by equation (2).

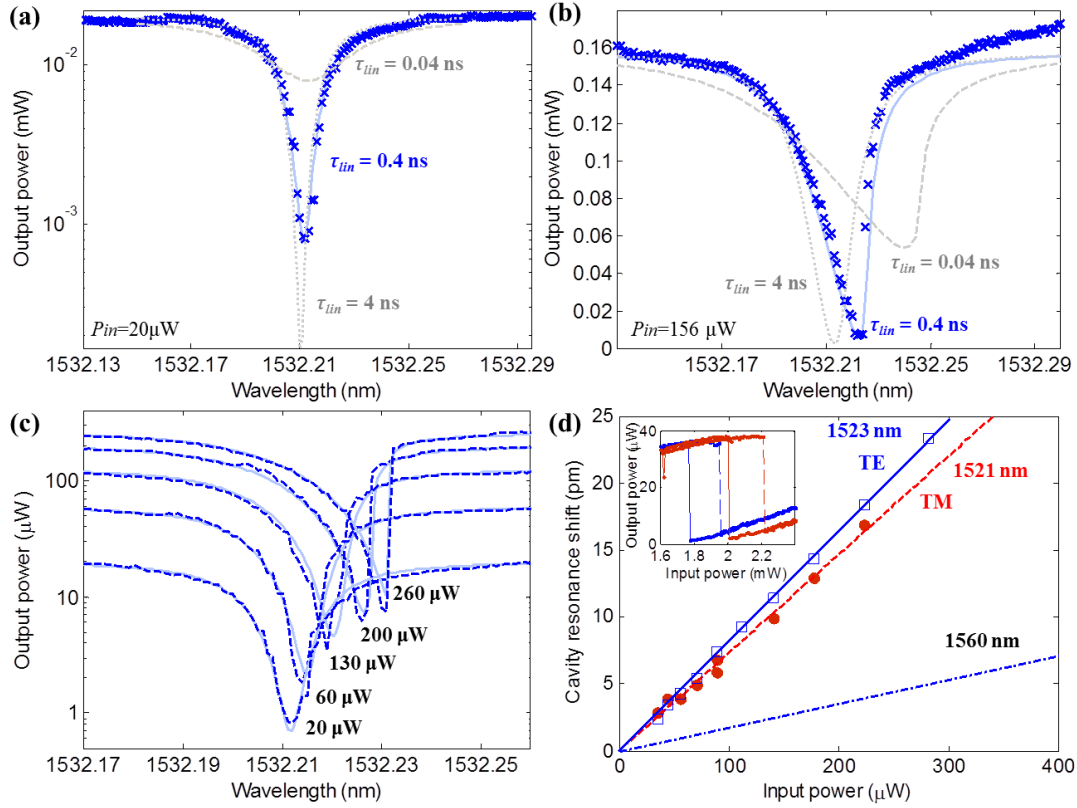

Fig. 3. Thermal nonlinearity in the microring resonances by molecular absorption (a) The transmission spectrum with input power at 20 μW for linear characterization. The dotted grey curve, solid blue curve, and the grey dashed curve are CMT simulation results with linear absorption rate of 1/4 ns, 1/0.4 ns and 1/0.04 ns respectively. The crosses are the experimental data. (b) The transmission spectrum with input power at 156 μW. (c) Optical transmission lineshape at different optical input powers (20, 60, 130, 200 and 260 μW). The dashed curves are experimental data and the solid curves are coupled mode theory simulations. (d) Cavity resonance shift versus the input power at defects absorption peak (80 pm/mW near 1520 nm) and away of the absorption peak (20 pm/mW near 1560 nm). Both TE (blue) and TM (red) polarizations are illustrated. Inset: measured hysteresis loop of the output versus input power near the 1523 nm resonance with TE polarization. The laser-resonance detunings are set at 33 and 34 pm for the blue and red lines respectively.

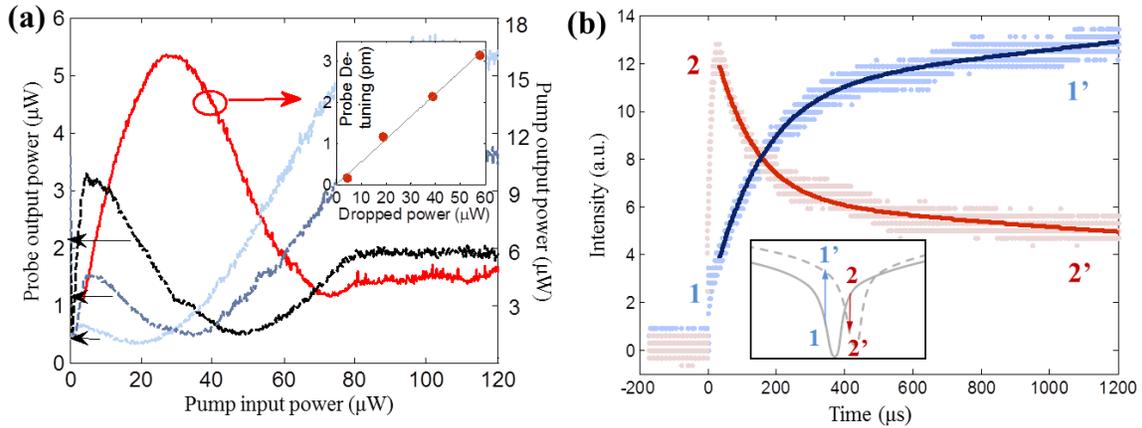

Fig. 4 Thermal optical bistability near the molecular vibration peak (a) Output power of pump and probe versus input pump power. The pump is 3.06 (red) to mode resonant at 1540.065 nm and the probes are set at 2.81 (light blue), 2.87 (navy), and 2.93 (black) to the mode resonant at 1542.962 nm Inset: The probe laser detuning to the resonance versus the pump power dropped into the resonator, driving the resonance to the correspondent probe laser wavelength. (b) Time domain self-heating dynamics to the step function input. The laser intensity step-function turns on to 1 mW. The laser-cavity detunings are -2 pm (blue) and 2 pm (red) respectively. The dots are experimental data and the lines are the exponential curve fitting. The lifetime is about 150 μs for both cases. Inset: Thermal switching dynamics for negative laser-cavity detuning (blue arrow) and positive detuning (red arrow), as the cold cavity resonance (solid grey line) red-shifted by thermal heating (dashed grey line).